\documentstyle[aps,prl,epsf,floats,axodraw]{revtex} 
\def\fig#1{fig.~{\ref{#1}}}
\def\figs#1#2{figs.~{\ref{#1}} and {\ref{#2}}}

\def\spa#1.#2{\left\langle#1\,#2\right\rangle}
\def\spb#1.#2{\left[#1\,#2\right]}

\def\pol{\varepsilon}

\def\Tr{\, {\rm Tr}}

\newbox\charbox
\newbox\slabox
\def\s#1{{      % Feynman slash
        \setbox\charbox=\hbox{$#1$}
        \setbox\slabox=\hbox{$/$}
        \dimen\charbox=\ht\slabox
        \advance\dimen\charbox by -\dp\slabox
        \advance\dimen\charbox by -\ht\charbox
        \advance\dimen\charbox by \dp\charbox
        \divide\dimen\charbox by 2
        \raise-\dimen\charbox\hbox to \wd\charbox{\hss/\hss}
        \llap{$#1$}
}}

\def\eqn#1{eq.~(\ref{#1})}

\def\qb{{\bar q}}
\def\Qb{{\bar Q}}
\def\ib{{\bar\imath}}
\draft

\begin{document}
\twocolumn[\hsize\textwidth\columnwidth\hsize\csname
@twocolumnfalse\endcsname 

hep-th/9912033 \hfill UCLA/99/TEP/55

\title{On the Coupling of Gravitons to Matter} 

\author{Z. Bern, A. De Freitas and H.L. Wong} 
\address{Department of Physics and Astronomy, UCLA, Los Angeles, CA
90095-1547 } 

\date{December, 1999}

\maketitle
             
\begin{abstract}
Using relationships between open and closed strings, we present a
construction of tree-level scattering amplitudes for gravitons
minimally coupled to matter in terms of gauge theory partial
amplitudes.  In particular, we present examples of amplitudes with
gravitons coupled to vectors or to a single fermion pair.  We also
present two examples with massive graviton exchange, as would arise in
the presence of large compact dimensions.  The gauge charges are
represented by flavors of dynamical scalars or fermions.  This also
leads to an unconventional decomposition of color and kinematics in
gauge theories.
\end{abstract}

\pacs{PACS numbers: 11.15.Bt, 11.25.-w, 11.55.-m, 11.25.Db, 95.30.Sf
\hspace{1.0cm} } 

\vskip2.0pc]

%% Section I %%%%%%%%%%%%%%%%%%%%%%%%%%%%%%%%%%%%%%%%%%%%%%%%%%%%%%%%%

\renewcommand{\thefootnote}{\arabic{footnote}}
\setcounter{footnote}{0}

%\section{Introduction}
%\label{sec-1}

Although gravity and gauge theories both contain local symmetries,
their dynamical behavior are rather different.  Nonetheless, there are
some rather striking relationships between the theories such as the
AdS/CFT correspondence~\cite{AdS} and the perturbative relationships
implied by the structure of closed string amplitudes in terms of open
string amplitudes~\cite{KLT}.  In this letter we demonstrate that at
least for the case of a flat background, there is an elegant and
powerful way to obtain the interaction of gravitons with matter, based
on the structure of heterotic strings. Our construction is valid for
the cases of gravitons coupled to vectors and to at most a single
fermion pair.  The cases of pure Einstein gravity and $N=8$
supergravity have already been discussed in
refs.~\cite{BGK,BDDPR,GravAllN}.  Recently, there has been a
resurgence of interest in perturbative gravity, especially in the
presence of large compact dimensions~\cite{Dimopoulos} and in AdS
spaces~\cite{AdS}, making it all the more timely to investigate the
perturbative structure of gravity theories.

The differences between gravitational and gauge interactions are well
known.  For example, quantum chromodynamics is renormalizable and
exhibits confinement. Neither property is shared by gravity.
Moreover, the structure of the Einstein-Hilbert and Yang-Mills
Lagrangians ${\cal L}_{\rm EH} = {2\over \kappa^2}\sqrt{-g} R$ and
${\cal L}_{\rm YM} = -{1\over 4} F_{\mu\nu}^a F^{\mu\nu a}$ are
rather different.  In particular, gravity has an infinite set of
contact interactions while non-abelian gauge theories have only up to
four-point contact interactions. 

At tree level, Kawai, Lewellen and Tye (KLT)~\cite{KLT} have, however,
expressed closed string amplitudes as sums of products of open string
amplitudes.  In the field theory limit this allows one to obtain
gravity tree amplitudes directly from gauge theory ones.  This is generally
simpler than using conventional gravity Feynman rules.  As discussed in
ref.~\cite{Grant}, by using non-linear field redefinitions and gauge
fixings, the Lagrangians of the two theories can be rearranged to make
the relations between the two theories more apparent. (See also
ref.~\cite{SiegelGrav}.)  With the analytic reconstruction method
described in refs.~\cite{Review,BDDPR,GravAllN} the loop amplitudes
can be obtained from the tree ones. In this way gravity can be
perturbatively quantized without reference to a Lagrangian.  This has
already been exploited to demonstrate that $D=11$ supergravity
diverges at two-loops in accordance with naive
expectations~\cite{BDDPR,Deser} and that $D=4$ $N=8$ supergravity
likely has the first divergence at five~\cite{BDDPR} instead of three
loops~\cite{Stelle}, as expected from superspace power counting.  In
order to perform a more general investigation of the divergence
structure of gravity theories, it would be useful to have more general
tree amplitudes besides those of pure gravity or supergravity.

Here we discuss amplitudes where gravitons are minimally coupled
to gauge theories.  Maximally helicity violating (MHV) $n$-point tree
amplitudes for pure gravity~\cite{BGK} and for gravitationally dressed
Parke-Taylor~\cite{ParkeTaylor} gluon amplitudes have been presented
in refs.~\cite{BGK,Selivanov}. Using the KLT relations, we extend
these results to the remaining MHV gravitationally dressed gluon 
amplitudes and to ones with a single fermion pair.
Our construction also applies to cases of non-maximal helicity
violation, but the MHV cases are the simplest to present.  We also
present four-point examples with a (massive) graviton exchange.

The KLT equations~\cite{KLT,BGK} for field theory gravity tree amplitudes
are, 
\begin{eqnarray}
&& {\cal M}_4 (1,2_,3_,4) =
     - i s_{12} A_4(1,2,3,4) 
         \tilde A_4(1,2,4,3)\,, \nonumber\\
&& {\cal M}_5(1,2,3,4,5) 
  =    \nonumber \\
  && \hskip .8 cm i\Bigl[ 
    s_{12} s_{34}  A_5(1,2,3,4,5)
             \tilde A_5(2,1,4,3,5) \label{KLTExamples} \\
& & \hskip 1.1 cm 
      + s_{13} s_{24} A_5(1,3,2,4,5) \, 
         \tilde A_5(3,1,4,2,5) \Bigr]\,, \nonumber
\end{eqnarray}
where $s_{ij} = (k_i+k_j)^2$. The $A_n$ and $\tilde A_n$ are
tree-level color-ordered gauge theory partial
amplitudes~\cite{ManganoReview}. Our notation is to suppress
polarizations, $\pol_i$, and momenta, $k_i$, leaving only the label
$i$ for each leg. Explicit forms of the KLT equations for an arbitrary
number of external legs may be found in ref.~\cite{GravAllN}.  We have
dropped the couplings, but for each gauge and gravitational
interaction appearing on the left-hand-side, factors of the couplings
$g$ and $\kappa/2$ (where $\kappa^2 = 32 \pi G_N$) are respectively
required.

Although the use of the KLT equations in the field theory limit does
not require a fully consistent string theory, some features of string
theory are retained.  For example, an oriented string will necessarily
contain antisymmetric tensor and dilaton states in its massless
spectrum.  When describing graviton exchanges, the KLT equations will
therefore also implicitly contain these states.  However, for the
cases dealt with in this letter, the antisymmetric tensor and dilaton
decouple, so that extra projections are not required to remove these
states, if unwanted.

In order to apply the KLT equations to the cases of gravity coupled to
matter, we must specify the factorization of each state into states
appearing solely in gauge theories.  In doing so, we must ensure that
the spin and charges for each state are the desired ones.  We
factorize each graviton into a product of vectors, each gluon into
either a product of a vector and a scalar or a product of two fermions
and each quark into a fermion and a scalar.  As motivated by the
structure of the heterotic string, the gauge charges of each physical
state are assigned as flavor charges carried by scalars or by
fermions.  At the linearized level these factorizations are trivial.
In a sense, we use the KLT equations to extend them to the fully
interacting theories.

\begin{figure}[ht]
\centerline{\epsfxsize 2.4 truein \epsfbox{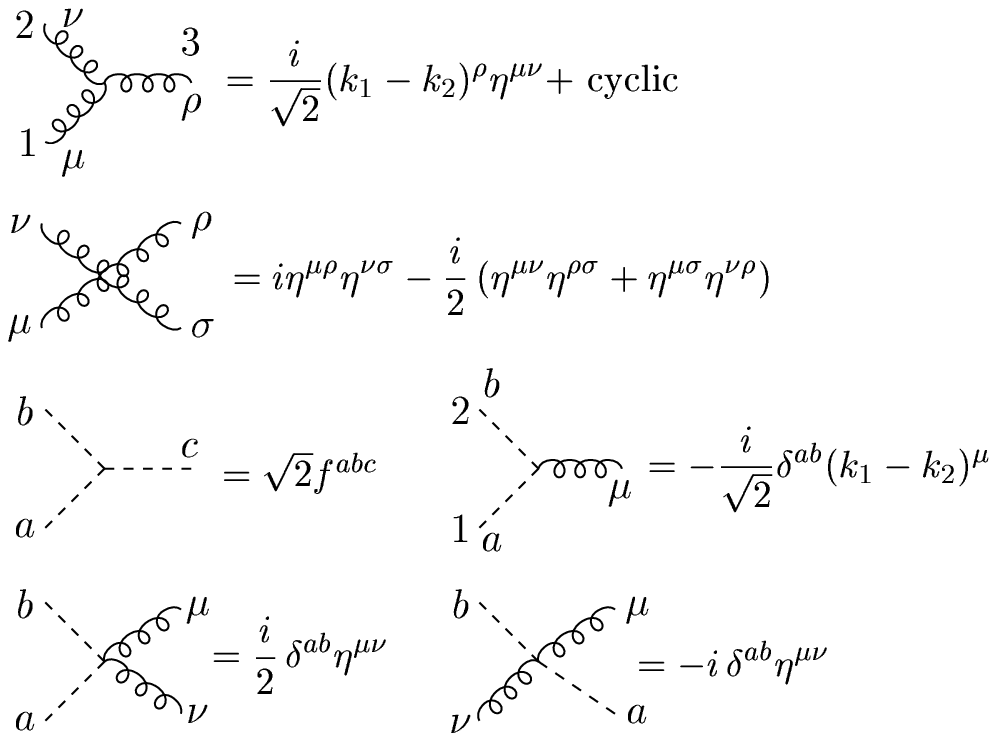}}
\caption[a]{\small The color-ordered Feynman vertices for obtaining,
via the KLT equations, gluon amplitudes dressed with gravitons.
Dashed lines represent scalars and curly ones vectors.}
\label{RulesFigure}
\end{figure}

As a specific example, consider $SU(N_c)$ gauge theory minimally
coupled to Einstein gravity. In \fig{RulesFigure} we present rules for
constructing, via the KLT equations, gluon amplitudes dressed by
minimally coupled gravitons; for this case, the rules give the partial
amplitudes $A_n$ and $\tilde A_n$ appearing in the KLT equations.
These rules are determined by requiring that the amplitudes with
gravitons carry the correct quantum numbers and dimensions.  As
discussed below, cases with a single fermion pair can be handled
similarly. (The signature of $\eta_{\mu\nu}$ is $(+,-,-,-)$.)  These
rules are similar to ordinary gauge theory Feynman rules except that
they are color ordered~\cite{ManganoReview}, i.e. the color group
theory factors have been stripped. In constructing the diagrams one
should maintain a clockwise ordering of the legs.  The group theory
factors appearing in the rule are associated with the flavors of
dynamical scalars. After substituting the $A_n$ and $\tilde A_n$ into
the KLT equations, on the left-hand-side the group theory factors are
reinterpreted as color factors.  We have displayed the vertices in the
commonly used Feynman gauge and have chosen all group theory factors
to be in the adjoint representation.  For other representations the
group theory factors should be altered to the appropriate one.

%%%%%%%%% FIGURE

It is instructive to first apply these rules to a pure gluon
amplitude.  The KLT equations will apply, since such amplitudes are
part of a theory coupling gravitons to gluons. In this case, the
construction is circular, since it starts with a gauge theory
amplitude only to wind up at the same amplitude. It will, however,
easily generalize to cases including gravitons.

With the standard color decomposition~\cite{ManganoReview}, the
four-gluon tree amplitude is expressed as,
\begin{eqnarray}
{\cal M}_4(1_g^{a_1},2_g^{a_2},3_g^{a_3},&& 4_g^{a_4}) 
= g^2 \sum_{\sigma } 
\Tr[T^{a_{\sigma(1)}} T^{a_{\sigma(2)}} T^{a_{\sigma(3)}}
  T^{a_{\sigma(4)}}] \nonumber\\
&& \times
  A_4(\sigma(1_g),\sigma(2_g),\sigma(3_g),\sigma(4_g)) \,,
\label{ColorDecomp}
\end{eqnarray}
where $\sigma$ runs over the set of non-cyclic permutations and the
$A_4$ are color-ordered partial amplitudes.  We denote gluon,
graviton, scalar and quark legs by the subscripts $g$, $h$, $s$ and
$q$. (Our fundamental representation color matrices are normalized by
$\Tr[T^a T^b]= \delta^{ab}$.)

\begin{figure}[ht]
\centerline{\epsfxsize 2. truein \epsfbox{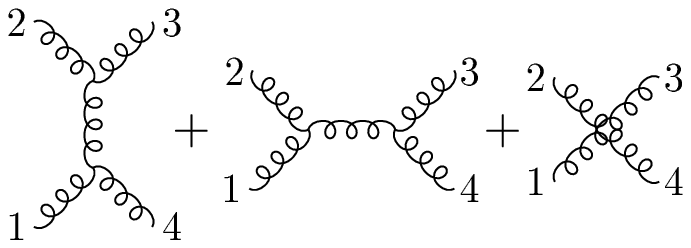}}
\caption[a]{\small The color-ordered Feynman diagrams contributing to 
$A(1_g,2_g,3_g,4_g)$.}
\label{GluonFigure}
\end{figure}

\begin{figure}[ht]
\vskip -.2 cm 
\centerline{\epsfxsize 1.4 truein \epsfbox{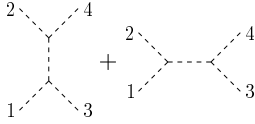}}
\caption[]{
\label{ScalarFigure}
\small The two diagrams contributing to the scalar partial amplitude
$\tilde A_4(1_s^{a_1},2_s^{a_2},4_s^{a_4},3_s^{a_3})$.  Each diagram is
dressed with flavor group theory factors.}
\end{figure}

By applying the KLT formula (\ref{KLTExamples}) we express the
four-gluon amplitude with only gluon exchanges as
\begin{eqnarray}
{\cal M}_4(1_g^{a_1},2_g^{a_2},3_g^{a_3},4^{a_4}_g)
 &=& - i g^2 s_{12} A_4(1_g,2_g,3_g,4_g)
 \nonumber \\ 
&& \hskip .4 cm \times
\tilde A_4(1_s^{a_1},2_s^{a_2},4_s^{a_4},3_s^{a_3}) \,,
\label{KLTGluons}
\end{eqnarray}
where the diagrams for obtaining $A_4(1_g,2_g,3_g,4_g)$ and
$\tilde A_4(1_s,2_s,4_s,3_s)$ are given in \figs{GluonFigure}{ScalarFigure}.
Note that in each of the two cases the labels on the diagrams follow
the cyclic ordering of the arguments of the amplitudes.  The value of
the first of these partial amplitudes is well known and may be found
in, for example, ref.~\cite{ManganoReview}.  Evaluating the diagrams
in \fig{ScalarFigure}, the four-scalar partial amplitude is,
\begin{eqnarray}
\tilde A_4(1_s^{a_1},2_s^{a_2},4_s^{a_4},3_s^{a_3}) =
 2 i {f^{a_1a_2b} f^{a_4a_3 b} \over  s_{12}} + 
 2 i {f^{a_2a_4b} f^{a_3a_1b} \over s_{13}} \,,  \nonumber
\end{eqnarray}
where $f^{abc} = - i \Tr\bigl( [T^a,T^b] T^c \bigr)/\sqrt{2}$ are the
group structure constants.  This type of decomposition of group theory
and kinematics is rather unconventional.  Nevertheless, it is
straightforward to verify that the amplitude in \eqn{KLTGluons} is
equal to that in \eqn{ColorDecomp}.  Similar decompositions hold for
any number of external legs, by applying $n$-point KLT equations.  In
this way both graviton and full gauge theory amplitudes satisfy KLT
equations, providing a natural way to obtain mixed amplitudes of
states appearing in gravity and in gauge theories.

%%%%%%% FIGURE

%%%%%%%%% FIGURE %%%%%%%%%%%%%%%%%%%%%%%%%%%%%%%

For example, we can express
${\cal M}_4(1_g,2_g,3_g,4_h)$, in terms of the gauge theory partial amplitudes
$A_4(1_g,2_g,3_g,4_g)$, whose diagrams are given in \fig{GluonFigure},
and the three-scalar one-vector partial amplitude $\tilde A_4
(1_s^{a_1},2_s^{a_2},4_g,3_s^{a_3})$ whose diagrams are the same as
those in \fig{ScalarFigure}, except that leg 4 is replaced
with a vector.

More generally, by applying the Feynman rules in \fig{RulesFigure} we
have obtained four-, five-, and six-point mixed amplitudes for gluons
and gravitons.  A sampling of these is, 
%
\iffalse
&& M_4(1_g^+,2_g^+, 3_h^-, 4_h^-) =  0\,, \nonumber \\
%
&& M_4(1_g^-, 2_g^+, 3_h^-, 4_h^+) = i g \Bigl({\kappa\over 2} \Bigr)
         {\spa1.4^4\spb1.3\spb3.2 
                     \over \spb3.1\spa1.2\spa3.4\spa3.1}\,, \nonumber\\
%
\fi
\begin{eqnarray}
&&M_5(1_g^+, 2_g^+, 3_g^+, 4_g^-, 5_h^-) = - i g^2 \Bigl({\kappa\over 2} \Bigr)
    {\spa4.5^4 \over \spa1.2 \spa2.3 \spa3.4 \spa4.1} \,, \nonumber \\
&&M_5(1_g^-,2_g^+, 3_g^+, 4_h^+, 5_h^-) = 
  i  g \Bigl({\kappa\over 2} \Bigr)^2
  {\spa1.5^4 \over \spa1.2 \spa2.3\spa3.1 } \nonumber \\
&& \hskip .8cm \times
 {1 \over \spa4.1 \spa4.5} 
   \Bigl[\spa2.1 \spa2.5 {\spb2.4 \over \spa2.4} + 
                            \spa3.1 \spa3.5 {\spb3.4 \over \spa3.4} \Bigr]\,,
  \nonumber \\
%
%&&M_5(1_g^+, 2_g^+, 3_g^+, 4_h^-, 5_h^-) = 0 \,, \nonumber \\
%
&&M_5(1_g^+, 2_g^+, 3_h^+, 4_h^-, 5_h^-) = 0 \,, \nonumber \\
&&M_5(1_g^-, 2_g^+, 3_h^+, 4_h^+, 5_h^-) = i \Bigl({\kappa\over 2} \Bigr)^3
    \spa1.5^4 \spa2.5^2\nonumber  \\
&& \hskip 5 cm \times
h(1,\{2,3,4\}, 5)
\,, \nonumber
\end{eqnarray}
for the coefficient of the color traces $\Tr[T^{a_i} \cdots T^{a_j}]$
which follow the ordering of gluon legs.  The complete amplitudes are
given by summing over non-cyclic permutations of the gluon legs after
multiplying by the color traces.  The function $h(a,\{P\},b)$
appearing in the last amplitude is the half-soft function defined in
ref.~\cite{GravAllN}.  Our convention for the `$\pm$' helicity
labeling is to take all legs to be outgoing.  In these amplitudes
graviton exchanges between gluons are not incorporated; such
contributions are suppressed by additional powers of the gravitational
coupling.  We have expressed the amplitudes in terms of $D=4$ spinor
inner products. (See e.g. ref.~\cite{ManganoReview}.)  These are
denoted by $\spa{i}.j = \langle i^- | j^+\rangle$ and $\spb{i}.j =
\langle i^+| j^-\rangle$, where $|i^{\pm}\rangle$ are massless Weyl
spinors of momentum $k_i$, labeled with the sign of the helicity.
They are antisymmetric, with norm $|\spa{i}.j| = |\spb{i}.j| =
\sqrt{s_{ij}}$.

Although $D=4$ spinor helicity is quite useful, the KLT equations do not
require its use, and are valid in arbitrary dimensions.  Likewise, the
conditions of maximal helicity violation are not essential, although
they considerably simplify amplitudes.  We have, for example, obtained
all the independent amplitudes with five gluons and one graviton using
the KLT equations and the six-gluon partial amplitudes given in
ref.~\cite{ManganoReview}.

The cases corresponding to gravitationally dressed Parke-Taylor
amplitudes match those previously computed by
Selivanov~\cite{Selivanov} using a generating function
method. Selivanov's $n$-point expression for the coefficient of
$\Tr[T^{a_1} T^{a_2} \cdots T^{a_m}]$ is
\begin{eqnarray}
&& M_n(1_g^-, 2_g^+, \ldots, i_g^-, \ldots, m_g^+, (m+1)_h^+,  \ldots,  n_h^+) 
     \nonumber\\
&& = i g^{m-2}\Bigl(-{\kappa\over 2}\Bigr)^{n-m} \hskip - .2 cm 
 {\spa1.i^4 \over \spa1.2 \spa2.3 \cdots \spa{m}.1} 
S(1,i, \{h^+\}, \{g^+\})\,, \nonumber 
\end{eqnarray}
where all legs have positive helicity except $1$ and $i$ and 
\begin{eqnarray}
S&&(i, j,\{h^+\}, \{g^+\}) = \biggl(\prod_{m\in \{h^+\}} 
   {d \over d a_m}\biggr) 
\nonumber\\
&& \times
 \prod_{l \in \{g^+\} } 
\exp\Bigl[\sum_{n_1 \in \{h^+\}} a_{n_1}
{\spa{l}.i \spa{l}.{j} \spb{l}.{n_1} \over 
 \spa{n_1}.i \spa{n_1}.j \spa{l}.{n_1} } \label{SelivanovFunc}\\
&& \times
\exp\Bigl[ \sum_{n_2 \in \{h^+\} \atop n_2 \not = n_1} \hskip -.1 cm a_{n_2}
 {\spa{n_1}.i \spa{n_1}.j \spb{n_1}.{n_2}  \over 
  \spa{n_2}.i \spa{n_2}.j
 \spa{n_1}.{n_2} } \exp\Bigl[\ldots \Bigr]\Bigr]\Bigr]
 \biggr|_{a_j = 0}\hskip - .3 cm  \,. \nonumber
\end{eqnarray}
In this factor legs $i$ and $j$ are the negative helicity ones while
the sets $\{h^+\}$ and $\{g^+\}$  represent
the positive helicity graviton and gluon legs.

The KLT equations immediately show that mixed gluon and graviton
amplitudes vanish if all legs have identical helicity or if there is a
single leg of opposite helicity, since the corresponding pure
gluon~\cite{ManganoReview} amplitudes vanish. We also extrapolate
our results to $n$-legs to obtain the remaining non-vanishing MHV
gluon amplitudes dressed with gravitons.  Doing so yields,
\begin{eqnarray}
\iffalse
&& M_n(1_g^+, \ldots, m_g^+, (m+1)_h^-, (m+2)_h^-, \nonumber\\
&& \hskip 4 cm  (m+3)_h^+, \ldots,  n_h^+) = 0\,, \nonumber \\
\fi
%
&& M_n(1_g^-, 2_g^+, 3_g^+, \ldots, m_g^+, (m+1)_h^-, (m+2)_h^+, 
                                                        \ldots, n_h^+) 
  \nonumber \\
&&  \hskip 1 cm 
 =  i g^{m-2}\Bigl(-{\kappa\over 2} \Bigr)^{n-m}
{\spa1.{\,m+1}^4 \over \spa1.2 \spa2.3 \cdots \spa{m}.1} \nonumber\\
&&  \hskip 2 cm \times
     S(1, m+1, \{h^+\} , \{g^+\} ) \,,
\label{AllNAnsatz}
\end{eqnarray}
for the coefficient of the color factor $\Tr[T^{a_1} T^{a_2} \cdots
T^{a_m}]$.  Furthermore, the amplitudes where the gluons are all of
identical helicity vanish, independent of the graviton helicity
configuration.  These amplitudes were constructed by demanding that
they satisfy the required factorization properties in all channels:
The amplitudes have no multi-particle poles, obey the collinear and
soft factorization for gluons given in ref.~\cite{ManganoReview} and
the ones for gravitons given in ref.~\cite{GravAllN}.  Moreover, they
have the correct phase singularity for a gluon becoming collinear with
a graviton.  Although this does not constitute a complete proof that
these results are valid for all $n$, in the past similar analytic
requirements have invariably led to the correct results.  (See
e.g. refs.~\cite{ParkeTaylor,AllNExamples}.)

The KLT equation (\ref{KLTExamples}) can also be directly applied to
the case of a single fermion pair, by using the vertex $i \gamma^\mu
/\sqrt{2}$ for the coupling of a fermion pair to a vector.  For
example, in \eqn{AllNAnsatz} the result of replacing legs 1 and 2 by
gluinos is given by multiplying by the overall factor $\spa2.{\,
m+1}/\spa1.{\, m+1}$.  This is consistent with the result obtained by
applying $N=1$ supersymmetry Ward identities~\cite{SWI}.  For
fundamental representation quarks one must also modify the associated
color factor to $(T^{a_3} \cdots T^{a_m})_{i_1}^{\ib_2}$.  Cases with
a single gravitino pair are similar.  On the other hand, cases with
multiple fermion pairs are more intricate. In particular, for the KLT
factorization to work, auxiliary rules for assigning global charges in
the color-ordered amplitudes appear to be necessary.

Although, we do not have a general formalism, it is interesting that a
KLT factorization can also be found for examples with a massive
graviton.  For example, the four-gluon amplitude with a
massive graviton exchange can be factorized as
\begin{eqnarray}
{\cal M}_4^{\rm ex}(&& 1_g^{-a_1}, 2_g^{+a_2}, 3_g^{-a_3}, 4_g^{+a_4}) 
                 \nonumber \\
&& = -i 
\biggl({\kappa \over 2} \biggr)^2
(s_{12} - m^2)\, A_4(1_\qb^{-a_1},2_q^{+a_2},3_\qb^{-a_3},4_q^{+a_4}) 
 \nonumber\\
&& \hskip 2 cm 
\times  \tilde A_4(1_\qb^-, 2_q^+, 4_q^+, 3_\qb^-)  \label{ggggExchange}\\
&& = -i \biggl({\kappa \over 2}\biggr)^2
\spb2.4^2 \spa1.3^2 
 \Biggl( {\delta^{a_1 a_2} \delta^{a_3 a_4} \over s_{12} - m^2} 
       + {\delta^{a_1 a_4} \delta^{a_2 a_3} \over s_{14} - m^2} \Biggr) \,.
 \nonumber
\end{eqnarray}
On the right-hand-side the color-ordered amplitudes are those for
quark scattering via massive vector boson exchanges.  (This
factorization works because of our particular assignment of $a_i$ charges.)
Similarly, the scattering of fundamental representation quarks by
gluons via massive graviton exchange is 
\begin{eqnarray}
{\cal M}_4^{\rm ex}(1_q^{-i_1},&& 2_\qb^{+\ib_2}, 3_g^{-a_3}, 4_g^{+a_4}) 
\nonumber\\
&&  = 
- i  \Bigl({\kappa \over 2}\Bigr)^2
  (s_{12} - m^2) A_4(1_s^{i_1}, 2_s^{\ib_2}, 3_Q^{-a_3}, 4_\Qb^{+a_4}) 
\nonumber \\
&& \hskip 1.5 cm \times
  \tilde A_4(1_q^-, 2_\qb^+, 4_\Qb^+, 3_Q^-) \label{qqggExchange} \\
&& = 
i \Bigl({\kappa \over 2}\Bigr)^2 \spa1.3^2 \spb4.2 \spb4.1 
 {\delta_{i_1}{}^{\ib_2}\delta^{a_3a_4} \over s_{12} - m^2} \,,
\nonumber
\end{eqnarray}
where $q$ and $Q$ are distinct massless fermion species.  For both amplitudes
(\ref{ggggExchange}) and (\ref{qqggExchange}) we have verified that
both sides of the equations are equal by explicitly computing the
amplitudes on the left-hand-side using conventional Feynman rules.

The examples (\ref{ggggExchange}) and (\ref{qqggExchange}) suggest
that it should be possible to obtain a general formalism for (massive)
graviton exchanges.  With multiple fermion pairs 
the antisymmetric tensor and dilaton will decouple 
only with extra projections.

%%%%%%%%%%%%%%%

It would be interesting to see if the relationships between
gravity and gauge theories discussed here can be extended to more
general field configurations.

We thank Lance Dixon and Maxim Perelstein for useful discussions.
This work was supported by the US Department of Energy under grant
DE-FG03-91ER40662.

%%%%%%%%%%%%%%%%%%%%%%%%%%%%%%%%%%%%%%%%%%%%%%%%%%%%%%%%%%%

\end{document}